\documentclass[journal]{IEEEtai}

\usepackage{cite}
\usepackage{multirow}
\usepackage[table,xcdraw]{xcolor}
\usepackage{graphicx}
\usepackage{amsmath,amssymb,amsfonts}
\usepackage{algorithmic}
\usepackage{graphicx}
\usepackage{textcomp}
\usepackage{xcolor}
\usepackage{subfigure}
\usepackage[graphicx]{realboxes}
\usepackage{cite}
\usepackage{amsmath,amssymb,amsfonts}
\usepackage{algorithmic}
\usepackage{graphicx}
\usepackage{textcomp}
\usepackage{xcolor}
\usepackage{multirow}
\usepackage{multicol}
\usepackage{bbding}
\usepackage{booktabs}

\def\BibTeX{{\rm B\kern-.05em{\sc i\kern-.025em b}\kern-.08em
    T\kern-.1667em\lower.7ex\hbox{E}\kern-.125emX}}
\begin{document}

\graphicspath{{figures/}}
    
\title{A Trustworthy AIoT-enabled Localization System via Federated Learning and Blockchain}

\author{%
  \IEEEauthorblockN{%
    Junfei~Wang,
    He~Huang,
    Jingze~Feng,
    Steven~Wong,
    Lihua~Xie,
    and~Jianfei~Yang
  }%
  
   \thanks{J. Wang, H. Huang, and L. Xie are with the School of Electrical and Electronics Engineering, Nanyang Technological University, Singapore. J. Yang is with the School of Mechanical and Aerospace Engineering and the School of Electrical and Electronics Engineering at Nanyang Technological University, Singapore. }
   \thanks{J. Feng is with Chainbase Inc.}
   \thanks{S. Wang is with PublicAI Inc.}
   
    \thanks{J. Yang is the corresponding author (yang0478@e.ntu.edu.sg).}
}


\markboth{}%
{Wang et al.: Decentralized Federated Learning for AIoT System}

\maketitle

\begin{abstract}
There is a significant demand for indoor localization technology in smart buildings, and the most promising solution in this field is using RF sensors and fingerprinting-based methods that employ machine learning models trained on crowd-sourced user data gathered from IoT devices. However, this raises security and privacy issues in practice. Some researchers propose to use federated learning to partially overcome privacy problems, but there still remain security concerns, e.g., single-point failure and malicious attacks. In this paper, we propose a framework named DFLoc to achieve precise 3D localization tasks while considering the following two security concerns. Particularly, we design a specialized blockchain to decentralize the framework by distributing the tasks such as model distribution and aggregation which are handled by a central server to all clients in most previous works, to address the issue of the single-point failure for a reliable and accurate indoor localization system. Moreover, we introduce an updated model verification mechanism within the blockchain to alleviate the concern of malicious node attacks. Experimental results substantiate the framework's capacity to deliver accurate 3D location predictions and its superior resistance to the impacts of single-point failure and malicious attacks when compared to conventional centralized federated learning systems.
\end{abstract}

\begin{IEEEImpStatement}
 This article introduces a resilient indoor localization solution grounded in a pioneering decentralized federated learning framework. To our knowledge, it marks the inaugural endeavor to delve into a decentralized approach for reliable indoor localization, validated through real-world data. Our method adeptly confronts the hurdles of single-point failure and malevolent attacks in current AIoT systems, showcasing enhanced robustness in comparison to prevailing solutions reliant on conventional machine learning or traditional federated learning systems. We anticipate that this research will safeguard the privacy of IoT users' information and build a trustworthy AI model for indoor localization in smart buildings.
\end{IEEEImpStatement}



\section{Introduction}
Nowadays, Location-Based Services (LBS) leverage the geographical data from IoT devices or users to provide tailored interaction services, significantly enhancing the IoT user experience, and are widely used in smart cities~\cite{zafari2019survey,zou2017non}. According to the application scenarios and requirements, it can be divided into outdoor and indoor localization~\cite{rizos2005trends}. Outdoor localization technologies, e.g., GPS, are adopted for driving and navigation~\cite{vo2015}. However, owing to complex indoor environmental factors and the absence of GPS signals, these outdoor positioning technologies cannot produce satisfactory results in the indoor environment, especially in complex multi-floor smart buildings, which motivates the research on indoor positioning technology. Indoor localization technology utilizes radio frequency (RF) signals such as WiFi, Bluetooth, and Ultra-Wideband (UWB) and locates the target using the received signal strength (RSS) that reflects the distance between the sensor and the user device~\cite{huang2023fusion,chen2019wifi}, leading to better performances in GPS-denied environments.

In the field of RSS-based indoor localization, one of the most prominent solutions is based on fingerprints and machine learning models due to their higher accuracy and adaptability. The indoor positioning algorithm of the fingerprint-based methods typically involves two primary phases: offline training and online localization~\cite{vo2015}. During the offline training phase, a collection of reference points is conducted by gathering RSS measurements from known locations within the indoor environment. These reference points are utilized to construct a fingerprinting database. Subsequently, the localization models employ machine learning or even deep learning algorithms to establish a mapping between the signal fingerprints and their corresponding locations. In the online localization phase, the established localization models can determine the device's position based on real-time RSS measurements.

\begin{figure}[tbp]
\centering 
\includegraphics[width=1\linewidth]{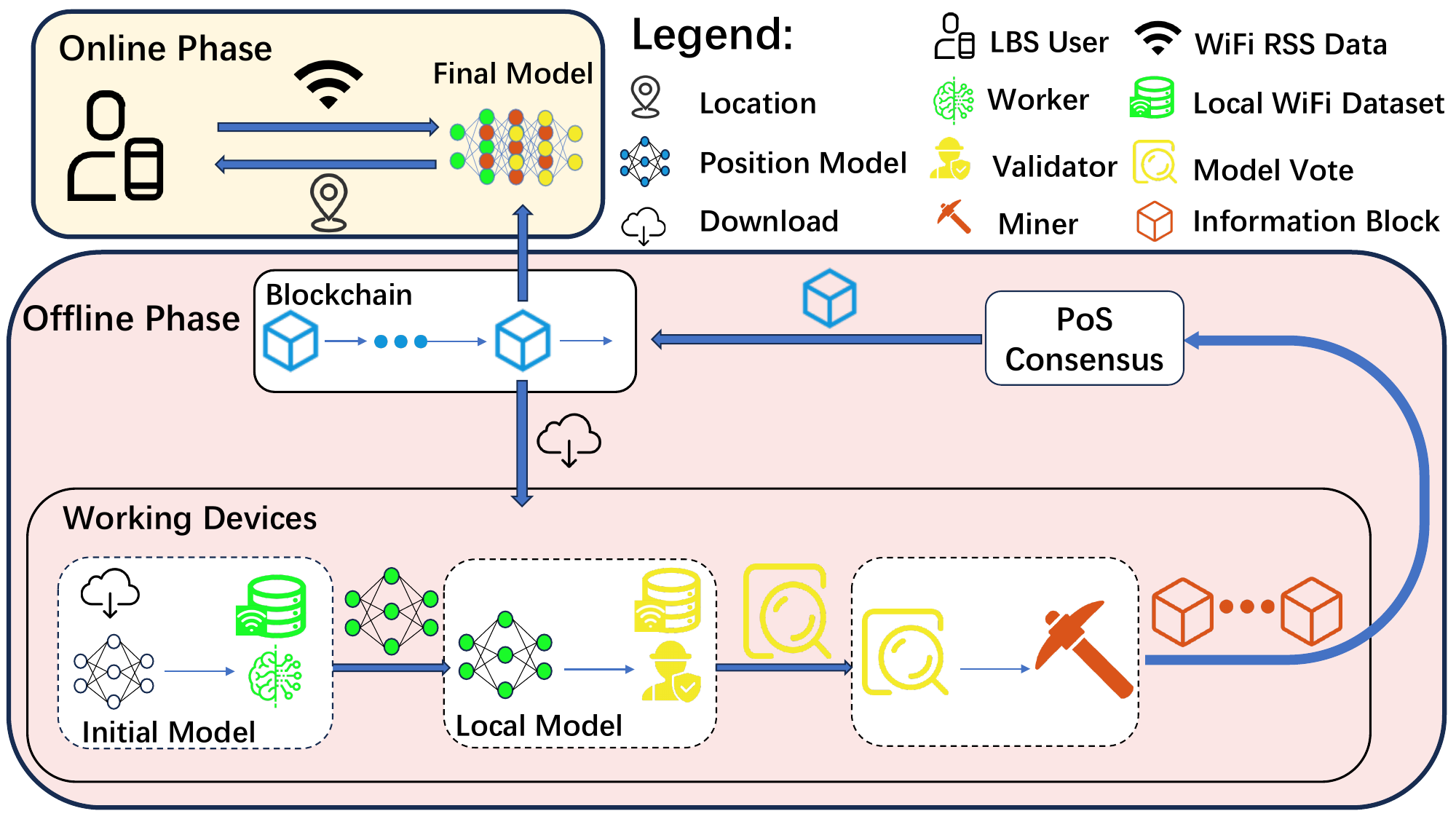}
\caption{DFLoc system overview. Clients handle tasks including model training, verification, and block mining, constituting a blockchain that can aggregate and distribute the global model, replacing the central server.}
\label{SystemOverview}
\end{figure}

In fingerprinting-based indoor localization systems, the training of the machine learning algorithm is conducted on a central server. The users of the localization service transmit their fingerprint data from the respective IoT devices to the central server. The central server engages in crowd-sourcing training data from users, trains a deep learning model, and provides indoor localization services, such as building floor classification (BFC) and latitude longitude regression (LLR)~\cite{santos2021crowdsourcing}. However, such a framework raises several concerns in practice. Firstly, sharing location-linked fingerprint data collected locally with a central entity can raise privacy issues. More and more users prefer not to share the data directly in a central server. Secondly, the training and inference of machine learning algorithms rely solely on a central server, posing a single-point failure, a common security issue in centralized learning systems; when the central server fails, the entire system becomes non-functional until it's restored~\cite{wang2021}. Thirdly, the system may be vulnerable to malicious attacks by dishonest clients through the injection of noisy data.

To address the privacy concern, Federated Learning (FL)~\cite{mcmahan2017} emerges as a promising solution, a machine learning technique that eliminates the need to transfer raw data from clients to the central server, mitigating privacy invasion. However, existing FL methods still suffer from security issues, e.g., single-point failure or malicious attacks. In fingerprinting-based indoor localization systems, single-point failure refers to the situation where the failure of the central server in the model training phase or implementation phase will cause the failure of the entire system. Malicious attacks entail the dispatch of nefarious clients by competitors or adversaries, disrupting the model training process by transmitting falsified or tainted data. On the one hand, as for single-point failure, opting for introducing decentralized technology to ease the over-reliance on the central server is a favorable solution. To this end, we introduce blockchain techniques due to their attributes of decentralization, traceability, and immutability. On the other hand, to counter malicious attacks, we design an update verification mechanism to differentiate between legitimate and malicious model updates, safeguarding our trained model from malicious alterations.

Specifically, we propose a Decentralized Federated Learning framework for the AIoT location system (DFLoc), as shown in Fig.\ref{SystemOverview}. 
The DFLoc is a kind of the fingerprint-based localization system, which consists of several clients. These clients are randomly assigned one of three roles (worker, validator, miner) and engage in model aggregation over multiple iterations to train a location prediction model. This model comprises a classification network, denoted as DFLoc-BFC, and a regression network, referred to as DFLoc-LLR. The roles randomly assigned to the clients include local workers responsible for actual training, update validators tasked with verifying update quality and providing votes, and blockchain miners who aggregate vote results and client stakes and upload them onto the blockchain. At the end of each round, all devices, regardless of their previous roles, collectively aggregate model parameters whose received positive votes exceed the negative votes. This ensures that models suspected of poor quality or tampering are excluded from the model aggregation process. In the experimental part, we use WiFi data to validate the effectiveness of our proposed DFLoc system, since WiFi is one of the most prevalent methods in fingerprint-based indoor localization systems~\cite{abbas2019wideep}.

Drawing upon the concepts above, this paper formulates and presents a novel decentralized federated location framework called DFLoc. The primary objective of DFLoc is to address the previously mentioned challenges and make the following contributions:
\begin{itemize}
    \item We aim to deal with the privacy and security issues in indoor localization by a decentralized federated learning framework. As far as we know, this is the first work that deals with both two issues simultaneously in RSS-based indoor localization systems.

    \item We revamp federated learning with blockchain algorithms to decentralize the system framework. This addresses the single-point failure problem caused by over-reliance on the central server.

    \item We design an update validation mechanism and integrate it into the blockchain to mitigate the impact of malicious nodes on the global indoor localization model.

    \item To explore the effectiveness of our proposed framework against single-point failure and malicious attacks, we take the traditional centralized federated learning (CFL) as the baseline to conduct some experiments on a real-world dataset of WiFi fingerprints.
\end{itemize}

The structure of this article is as follows: Section~\ref{II} reviews related work. Section~\ref{III} provides an overview of the fingerprinting-based indoor localization problem first, then details our proposed framework, DFLoc, including its operation and the DFLoc validator mechanism. Section~\ref{IV} presents the performance evaluation. Finally, Section~\ref{V} offers the conclusion for the entire work.

\section{Related Works} \label{II}

Two principal concerns arise in the current research on 3D Indoor Location Systems based on the federated learning. The first concern pertains to the challenge of accurate Indoor 3D localization, while the second revolves around the reliability of federated learning.

\subsection{Fingerprint-based Indoor Localization}
Within indoor environments, fingerprinting-based localization stands out as a favored choice, due to its ease of realization and remarkable accuracy. Implementing a fingerprint-based approach involves deploying several wireless sensors capable of acquiring RSS and creating a fingerprinting database, which consists of a set of reference points paired with the corresponding RSS values of the selected signal type for each reference point. Mainstream electronic devices used in daily life, such as laptops and IoT devices, all come with built-in WiFi capability~\cite{zafari2019survey}. Concurrently, existing WiFi access points can be used as reference points for signal collection~\cite{kumar2014accurate}. These factors lead to WiFi as the most prevalent sensing technique for RSS fingerprint-based indoor localization. Some scholars apply centralized learning techniques as a means to address the challenge of WiFi fingerprinting-based indoor localization~\cite{zhang2016,khatab2017,kim2018,huang2022varifi}. However, all these centralized learning approaches have no longer been the preferred choices when considering personal information privacy and dynamic environment. On the contrary, FL is a more suitable solution for this problem due to the faster direct responses to newly measured data and the lower privacy loss~\cite{nguyen2021,wu2022,park2022,li2020p}. 

Some FL methods are applied to the WiFi fingerprinting-based localization field. Nonetheless, these approaches encounter certain challenges. For instance, in their respective studies, \cite{wu2022} and~\cite{park2022} focus solely on the issue of LLR within a limited 2D environment. While \cite{li2020p} tackles both BFC and LLR problems within a 3D environment, it simplifies the latter regression task to a discrete classification task, resulting in coarser location estimates. In response to the imperative of precise 3D indoor localization estimation, \cite{gao2022federated} introduce FedLoc3D, a framework incorporating both classification and regression networks to resolve floor classification and longitude-latitude regression problems, respectively. The fusion of results from these two sub-networks yields accurate 3D coordinates. However, all of these works lack consideration of the concerns of single-point failure and malicious attacks. Inspired by their work, we design our sub-networks to enhance our approach and solve the security concerns mentioned above.

We draw inspiration from the network structure depicted in their work to enhance our approach and consider the two security concerns mentioned above to design a robust and accurate 3D coordinate indoor localization system.

\subsection{Federated Learning and Blockchain in Localization}

Federated learning, as introduced by Google~\cite{mcmahan2017}, represents an innovative approach designed to address privacy concerns within the realm of machine learning. This framework ensures the preservation of data privacy by permitting clients to maintain their private datasets without necessitating data transfers~\cite{zhang2020}. Nevertheless, the implementation of federated learning may bring two significant problems, commonly referred to as single-point failure and malicious attack problems. On one hand, the single-point failure stems from an over-reliance on central servers for critical functions, such as local training aggregation. In the event of a central server malfunction, the entire system becomes vulnerable to complete disruption~\cite{issa2023}. On the other hand, malicious attacks can occur when certain clients deviate from prescribed protocols, introducing noise or injecting compromised parameters/models during the local model training process~\cite{wang2021}. These two challenges bring innegligible threats to the reliability of federated learning systems and raise significant attention in both research and industrial fields.

To handle the issues mentioned above, researchers explore the integration of several decentralized techniques and FL to strengthen the traditional CFL architecture. Blockchain, a decentralized ledger technology to reach a consensus on the shared transaction data, emerges as a promising solution, whose attributes of decentralization, traceability, and immutability make it a compelling choice for addressing the aforementioned challenges. As a result of the integration of blockchain and FL, Blockchain-based federated learning (BCFL)~\cite{li2022} can mitigate the single-point failure and malicious attacks~\cite{ramanan2020,kim2019,chen2021}. However, there is no reported research work proposed to solve the security issue in localization systems.

\section{Methodology}\label{III}
This section will describe the problem definition of the fingerprinting-based indoor localization in Section~\ref{problem}, the proposed framework DFLoc in Section~\ref{operation}, and the DFLoc validator mechanism in Section~\ref{ValidatorMechanism}.

\begin{figure}[!t]
\centering 
\includegraphics[width=\linewidth]{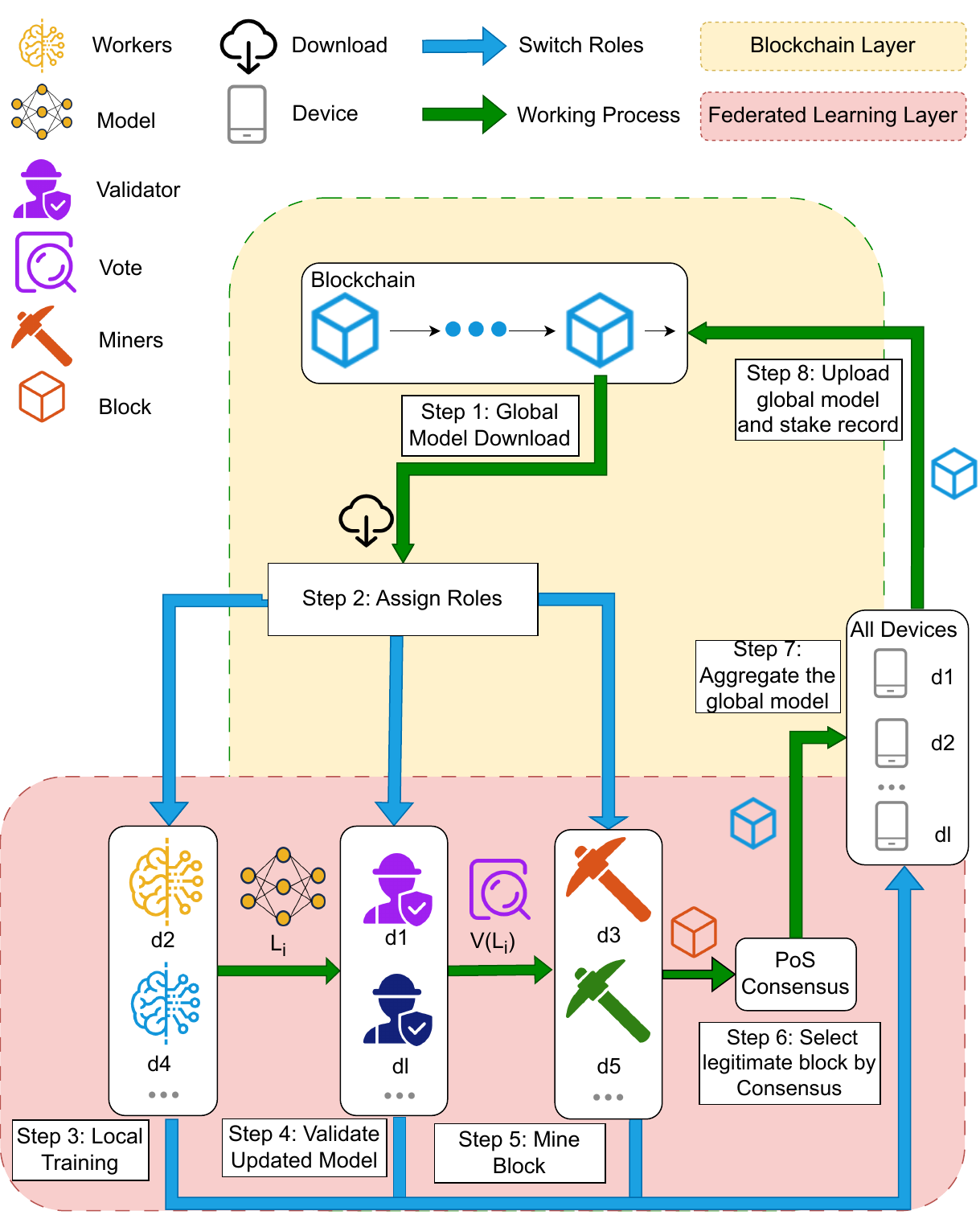}
\caption{Operations of DFLoc. In each round of the DFLoc learning phase, after downloading the global model, each client is assigned a specific role and completes the corresponding task. Subsequently, all clients aggregate the global model and update it with the stake record onto the blockchain.}
\label{Method}
\end{figure}

\subsection{Problem Definition} \label{problem}

In the fingerprinting-based indoor localization field, users seek to determine their precise locations by requesting the LBS from the server where a deep learning model is trained in advance to provide location estimate based on the fingerprints of the RF signal~\cite{gao2022federated}. Suppose that we have a dataset comprising RF fingerprints labeled with locations, denoted as $D=\{x_i, y_i\}|_{i \in \mathcal{N}}$, where $\mathcal{N} =\{1,2,...,n\}$ denotes the data sample indices, 
$x_i$ and $y_i$ represent the measured RSS vector and the location including discrete building-floor indexes ($BL_i$ and $FL_i$) and continuous latitude and longitude values ($LA_i$ and $LO_i$) at some reference point, respectively.

It is worth noting that, although our method is implemented using WiFi RSS data in this work, the methodology could seamlessly accommodate various RF signal types, such as UWB and Bluetooth. The primary goal of fingerprinting-based localization resides in the acquisition of a location prediction model denoted as $f (\cdot; \omega)$, characterized by parameters $\omega$, to minimize the following average loss: 

\begin{equation}
     \mathop{\min}_{\omega \in \mathbb{R}^d} \epsilon(\omega)
\end{equation}
Given the dataset $D$ that contains $n$ data samples, the localization loss is further defined as:
\begin{equation}
    \epsilon(\omega) \triangleq \frac{1}{n} \sum_{i \in D}
    l(x_i;y_i;\omega)
\end{equation}
where $l(x_i;y_i;\omega)$ denotes a loss function that characterizes the error between $i$-th location prediction $\hat{y}=f(x_i;\omega)$ and its ground truth value $y_i$.

In traditional indoor localization systems that apply the centralized machine learning technologies~\cite{nguyen2021,wu2022,park2022,li2020p}, all user data needs to be stored on the central server for training. When data is transmitted or stored on the server, risks of exposing the privacy of the data producers arise. Furthermore, this architecture's heavy reliance on the central server may bring a single-point failure problem, which means if the central server fails, the entire system would lose its functionality. Moreover, it's not reasonable to assume that all the clients are honest. The performance of traditional systems is very vulnerable to malicious attacks like uploading tampered data or local models. Thus, we propose DFLoc to solve the problems mentioned above.

\subsection{Decentralized Localization System} \label{operation}

As shown in Fig.~\ref{Method}, the operation of the DFLoc framework is maintained by a group of devices,  denoted as $\mathcal{D}=\{d_1, d_2, ..., d_l\}$, across $R$ communication rounds. At the onset of the $i$-th round, all devices are required to retrieve the current global model $G_{i-1}$ from the blockchain. Furthermore, in each epoch, every device is assigned randomly to one of three temporary roles: worker, validator, or miner. Consequently, these devices are transiently grouped into three distinct categories, denoted as $\mathcal{W}$, $\mathcal{V}$, $\mathcal{M}$, and $|\mathcal{D}| = |\mathcal{W}|+|\mathcal{V}|+|\mathcal{M}|$.

Each worker $w \in \mathcal{W}$ proceeds to train its local updated model $L_i^w$ using its respective training dataset $D_{tr}^w$. The reward for worker $w$ is calculated as $r_i^w=e^w_i \times |D_{tr}^w| \times r$, where $r$ is a unit reward. $e^w_i$ and $|D_{tr}^w|$ denote the number of local training epochs and the size of the data samples within $D_{tr}^w$ at the $i$-th communication round, respectively. It is noticeable that not all workers provide efficient information for the blockchain. Thus, a validation mechanism, which will be detailed in subsection ~\ref{ValidatorMechanism}, is designed to distinguish valid workers, whose positive votes exceed negative votes evaluated by the following validators. Once a worker is seen as invalid, the relative reward will be zero. As a result, the transaction data from worker $w$ can be expressed as $Tw_i^w=\{L_i^w, r_i^w\}$, which is further signed by the private signature and sent to the subsequent validators. 




Subsequently, each validator $v \in \mathcal{V}$ examines worker transactions $Tw_i^w$, one at a time, in a sequential manner until all transactions are verified. Different validators should share the worker transactions they received with peers, ensuring that each validator receives all worker transactions for the current communication round. Subsequently, $v$ discards the transaction data without a digital signature, proceeds to extract $L_i^w$ from $Tw_i^w$ with a digital signature and evaluate its validity using the DFLoc validator mechanism. Afterward, $v$ issues either a positive or negative vote, denoted as $V^v(L_i^w)$, based on the outcome of the validation process. Following this, $v$ computes the reward, denoted as $r^v(L_i^w)$. Analogous to workers, the reward allocated to $v$ is proportionate to the size of its training dataset $D_{tr}^v$. As a result, the total reward accruing to $v$ during $i$-th communication round is the sum of all $r^v(L_i^w)$, and this can be expressed as follows

\begin{gather*}
   r^v(L_i^w)= |D_{tr}^v| \times r \\
   r_i^v = \sum_w r^v(L_i^w) = |\mathcal{W}| \times |D_{tr}^v| \times r.
\end{gather*}

Then, $v$ should encapsulate $Tw_i^w$, $V^v(L_i^w)$ and $r^v(L_i^w)$ together to obtain a validator transaction $Tv^v(L_i^w)=\{ Tw_i^w, V^v(L_i^w), r^v(L_i^w)\}$, which is then dispatched to the miner which is randomly associated with $v$ after signed by $v$'s private key. Each miner broadcasts to its peers, ensuring the dissemination of all validator transactions among all miners. By doing so, each local updated model corresponds to $|\mathcal{V}|$ validator transactions, and each miner will totally receive $|\mathcal{W}| \times |\mathcal{V}|$ validator transactions.

Each miner $m \in \mathcal{M}$ verifies the signature of each validator transaction $Tv^v(L_i^w)$. Upon successful signature verification, $m$ proceeds to extract $Tw_i^w$, $V^v(L_i^w)$, and $r^v(L_i^w)$ from the transaction to summarize the vote results of each local updated model $L_i^w$ (denoted as $SV^m(L_i^w)$) and the total reward of each validator (denoted as $r_i^v$) in the $i$-th communication round. Meanwhile, $m$ gets rewards for summarizing the results, denoted as $r_i^m$, which can be expressed as
\begin{equation}
    r_i^m = |\mathcal{W}| \times |\mathcal{V}| \times r .
\end{equation}

Afterward, each miner $m$ undertakes the consolidation of summarized results, including all local updated models $L_i^w$ and corresponding summarized votes $SV^m(L_i^w)$, all worker rewards $r_i^w$, all validator rewards $r_i^v$, and its own reward $r_i^m$, to assemble a candidate block denoted as $B_i^m$. Subsequently, miner $m$ engages in the mining process specific to the Proof of Stake (PoS) consensus which involves hashing the complete content of the block and signing the hash by using its private key. Upon $B^m_i$ being mined, miner $m$ propagates the mined block to all the other miners in the network. 

Following this dissemination, each miner $m$ leverages the stake information recorded on its blockchain to identify and select the block generated by the miner with the highest stake among the set $M$ as the legitimate block $B_i$. Only this legitimate block is deemed suitable for extracting the records of rewards $r_i^w$, $r_i^v$, and $r_i^m$ and summarized votes $SV^m(L_i^w)$, along with their corresponding model updates $L_i^w$. Furthermore, each $m$  broadcast this legitimate block to its peers and associated workers $w$ and validators $v$ to ensure that all devices receive this legitimate block $B_i$.

Finally, each device, denoted as $d$, regardless of its previous role, is tasked with two pivotal responsibilities to finish the $i$-th communication round. Firstly, it engages in the aggregation of locally updated models in legitimate block $B_i$ whose count of positive votes is not less than that of negative votes. This aggregation process yields a new global model $G_i$. Secondly, each device undertakes the task of updating the stake information by accumulating the stake records, and uploading the global model $G_i$ and updated stake records onto each individual blockchain copy to finish the $i$-th communication round. The entire iteration process continues over multiple rounds until the model reaches the desired performance.

\subsection{DFLoc Validator Mechanism} \label{ValidatorMechanism}

To enable the DFLoc system to withstand the impact of malicious attacks, a mechanism is devised to identify whether the updated local model contains malicious alterations. This gives rise to the DFLoc Validator mechanism.

In the $i$-th communication round, a validator $v$ typically evaluates the quality of the update model $L_i^w$ by comparing its testing localization accuracy $A^w(L^w_i)$ against that of a single-epoch trained local model, denoted as $A^w(L^w_i(1))$, on the worker's test dataset $D_{te}^w$, as suggested by~\cite{chen2021}. If noise distorts $L^w_i$, $A^w(L^w_i)$ will differ, leading to a decline in accuracy compared to $A^w(L^w_i(1))$. Conversely, unaltered $L^w_i$ yields minimal differences between $A^w(L^w_i)$ and $A^w(L^w_i(1))$. Notably, $v$ lacks access to $D_{te}^w$, so it cannot directly obtain the value pair $\{A^w(L^w_i(1)), A^w(L^w_i)\}$.

A viable solution to address this issue involves validator $v$ initially conducting a single-epoch of local learning by using global model $G_{i-1}$ and its train dataset $D_{tr}^v$ to obtain a local update model $L^v_i(1)$, and computing the performance of $L^v_i(1)$ and $L^w_i$ under $v$'s test dataset $D_{te}^v$, denoted as $A^v(L^v_i(1))$ and $A^v(L^w_i)$, respectively. Subsequently, they serve as the proxy evaluation for $A^w(L^w_i(1))$ and $A^w(L^w_i)$~\cite{chen2021}.

In BFC, validator $v$ evaluates the potential distortion of $L^w_i$ by calculating the validation accuracy difference, denoted as $d^v_{BFC} = A^v(L^v_i(1))-A^v(L^w_i)$, and comparing it to a validator-threshold value, $T_{BFC}^v$. The hypothesis behind this is that when a validator $v$ produces $A^v(L^v_i(1))$, the value of $d^v_{BFC}$ is expected to differ between the $L^w_i$ sent by a legitimate $w$ and that by a malicious $w$~\cite{chen2021}. If $d^v_{BFC} \geq T_{BFC}^v$, indicating that the accuracy drop exceeds $v$'s tolerance threshold, validator $v$ assigns a negative vote to $L^w_i$. Otherwise, $v$ issues a positive vote. Finally, we determine $V^v(L_i^w)$ through this method, which is expressed as below

\begin{equation}
V^v(L_i^w)=\left\{
\begin{array}{rcl}
1 &  {if \quad d^v_{BFC} \leq T_{BFC}^v}\\
-1 &  {if \quad d^v_{BFC} > T_{BFC}^v}\\
\end{array} \right. .
\end{equation}

In the LLR part, validator $v$ calculates the validation loss ratio, denoted as $r^v_{LLR} = A^v(L_i^v(1))/A^v(L_i^w)$, following the computation of $A^v(L^v_i(1))$ and $A^v(L^w_i)$. This ratio is then compared to a validator-threshold value, $T_{LLR}^v$, to assess the potential distortion of $L^w_i$. The premise here is that the value of $r^v_{LLR}$ will exhibit disparities between legitimate $L^w_i$ and malicious $L^w_i$~\cite{chen2021}. Similarly, we determine $V^v(L^w_i)$ using this methodology, which is expressed as below

\begin{equation}
V^v(L_i^w)=\left\{
\begin{array}{rcl}
1 &  {if  \quad r^v_{LLR} \geq T_{LLR}^v}\\
-1 &  {if \quad r^v_{LLR} < T_{LLR}^v}\\
\end{array} \right. .
\end{equation}

\section{Experiments and Simulation} \label{IV}

In this section, we first describe the experimental setting and then evaluate the efficiency of the proposed system when confronted with malicious attacks and single-point failure, as well as its performance within a 3D environment.

\subsection{Experimental Setting}

\subsubsection{Dataset}

Without loss of generality, our experiments are conducted on the most commonly used dataset in indoor localization, which is called UJIIndoorLoc~\cite{torres2014ujiindoorloc}. This dataset encompasses $21,049$ labeled WiFi fingerprint data points collected from three campus buildings, each having four or five floors. Each data point is characterized by $520$ RSS values and the associated location information. These RSS values are derived from $520$ wireless access points distributed throughout these buildings. The data elements comprise building and floor information, as well as latitude and longitude coordinates (in meters using UTM from WGS84). Since we aim to deal with privacy and security issues, we avoid the potential variations caused by environmental dynamics by selecting $20\%$ of the original training data to create a new testing set, as described in~\cite{gao2022federated}. The remaining training data is partitioned into $l$ ($l = |\mathcal{D}|$) data segments, assigned to individual clients for simulating the FL system. To tackle the classification and regression challenges in complex 3D scenes, such as smart buildings, we employ a dual-network approach. This approach involves the training of two distinct networks: DFLoc-BFC and DFLoc-LLR, each meticulously designed to tackle their respective specialized challenges. In this section, we provide a comprehensive overview of the structures of both networks.

\subsubsection{Network Architecture} To solve the BFC problem, we design a classification model called DFLoc-BFC, which takes an RSS measurements vector $x_i$ as input. Every element in this vector corresponds to the measured RSS value of a distinct access point. In the extensive 3D space, the input vector can become lengthy. For instance, in the UJIIndoorLoc dataset, every WiFi fingerprint record contains a $520$-element RSS vector from $520$ access points within the 3D space. Additionally, input vectors can be sparse due to limited wireless sensing coverage, resulting in many invalid values where certain access points are undetectable at specific locations. Therefore, we integrate a fully connected feature extractor into the input layer to reduce data dimensions and extract features. Furthermore, to capture label similarities among input vectors from the same building and floor, we employ 1D convolutional layers to extract key features, as proposed in~\cite{gao2022federated}. Meanwhile, a fully connected layer following the 1D convolution is designed to convert the vector to $8$ bits one-hot encoding which encompasses a $3$-bit segment designated for the explicit denotation of the building index, with the remaining $5$ bits allocated for the denotation of the floor index. We set two $argmax(\cdot)$ functions after these two groups of bits to output the final classification results of the DFLoc-BFC network $\hat{y}_{BFC}=(BL_i, FL_i)$.

As mentioned above, input vectors collected from a considerable 3D space covered by many access points may be very lengthy and sparse, so DFLoc-LLR also requires a lightweight feature extractor to implement data dimensionality reduction and feature extraction functions. Then, the following linear layers convert the output of the extractor into a vector of two real numbers $\hat{y}_{LLR}=(LA_i, LO_i)$, which denote the longitude and latitude of the reference point estimated by the DFLoc-LLR network, respectively. Finally, we combine the outputs of the two networks together to form the final 3D coordinate output and evaluate the 3D localization performance of DFLoc in the experiments in Subsection~\ref{3D_experiment}.

\subsubsection{Implementation Details}
All experiments are conducted on a computer with two NVIDIA 2080ti GPUs, an Intel(R) Core(TM) i9-10900K CPU @ 3.70GHz, and $125$ gigabytes of available RAM. Our DFLoc framework is trained under the following conditions: We utilize a total of $20$ client devices, consisting of $12$ workers, $5$ validators, and $3$ miners. Given the substantial differences in the required epochs for convergence between the two networks, we conduct separate training for each network.
In DFLoc-BFC, we adopt the BCELoss as our chosen loss function, employing the Adam optimizer~\cite{kingma2014adam}. The selected learning rate is set at $0.001$, with a batch size of $100$. There are $100$ communication rounds with each containing $10$ local epochs. The validator-threshold parameter is configured to $T_{BFC}^v=0.1$. In the configuration of DFLoc-LLR, the optimizer, batch size, and local epoch count remain unchanged. However, we switch the loss function to L1Loss, adjust the learning rate to $0.002$, and extend the number of communication rounds to $500$. Furthermore, the validator-threshold parameter is set to $T_{LLR}^v=0.9$.
Throughout our experiments, we employ a conventional centralized federated learning system as the baseline, abbreviated as CFL. Our proposed DFLoc framework is referred to as DFL in the following experiments and analysis for ease of reference.

\subsection{Evaluation on the effect against malicious attacks}

This section evaluates the proposed framework's effectiveness against malicious attacks in both LLR and BFC components. In the CFL system, malicious devices inject Gaussian noise (zero mean vector and identical covariance matrix) into model parameters which are uploaded during training iterations, to emulate the malicious attacks. In DFLoc, malicious workers also distort local model updates with similar noise.

\subsubsection{Evaluation on the effect against malicious attacks on DFLoc-LLR} \label{EofLLRMa}

\begin{figure}
    \centering
    \includegraphics[width=\linewidth]{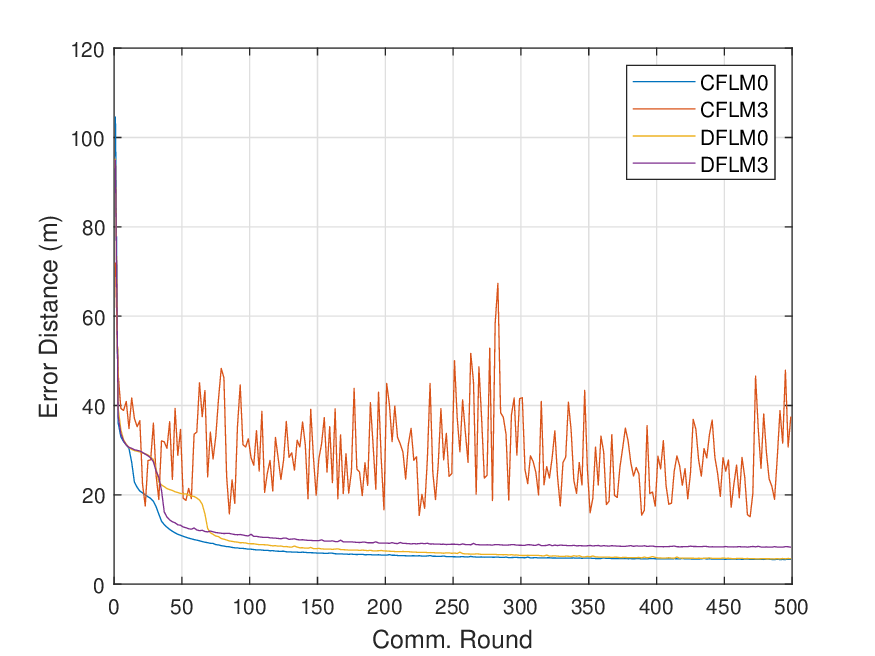}
    \caption{Training process of CFL and DFLoc-LLR under malicious attacks.}
    \label{test5}
\end{figure}

\begin{figure}
    \centering
    \includegraphics[width=\linewidth]{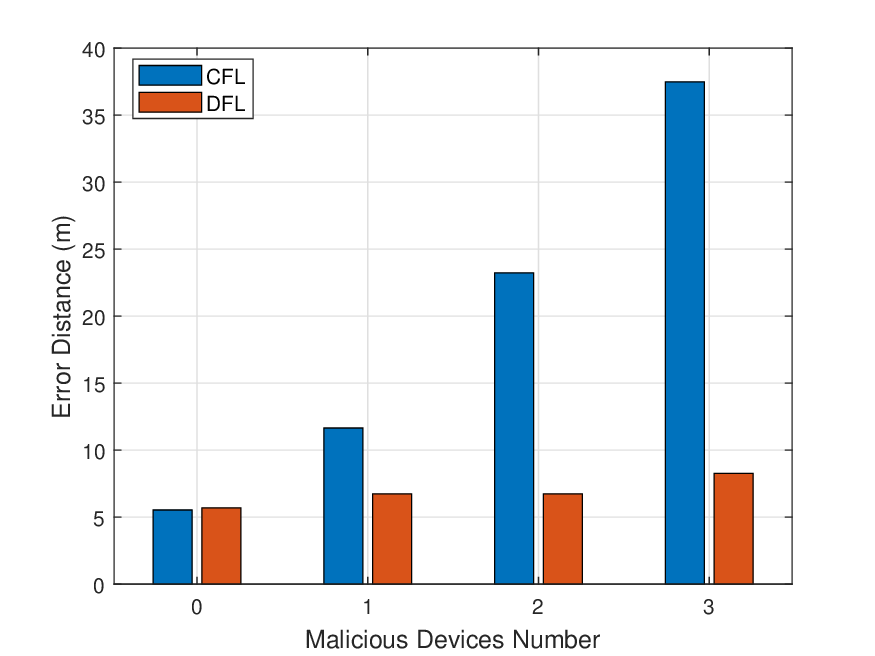}
    \caption{Effect of malicious attacks on CFL and DFLoc-LLR.}
    \label{test6}
\end{figure}

We conduct an experiment about four schemes: CFL without malicious devices (CFLM0), CFL with $3$ malicious devices (CFLM3), and two analogous schemes for DFLoc (DFLM0 and DFLM3). In Fig.~\ref{test5}, the orange curve illustrates CFLM3's error distance, highlighting its sensitivity to noisy model updates from malicious devices, resulting in unstable convergence.
The comparison between the unaffected blue curve (CFLM0) and the yellow curve (DFLM0) demonstrates similar performance, indicating the functionality of the FL components. Moreover, the purple curve (DFLM3) steadily converges to a relatively low error distance, albeit slightly higher than the blue and yellow curves. This is reasonable, considering that the training sets of malicious devices, which comprise $15\%$ of the total, are never effectively incorporated into the learning process. 

In Fig.~\ref{test6}, we present four pairs of schemes, each comprising blue (CFL) and orange (DFLoc) bars, reflecting different numbers of malicious devices. Irrespective of the varying count of malicious nodes, DFLoc maintains a consistently low error distance (below $8.5$ m), while CFL's performance is notably affected. This underscores the significant impact of malicious attacks on CFL, particularly with a larger number of malicious nodes. In contrast, DFLoc always demonstrates robust resistance to malicious attacks in this work. When malicious nodes constitute $15\%$ of the total, CFL's error distance is four times greater than that of DFLoc.

\begin{figure}
    \centering
    \includegraphics[width=\linewidth]{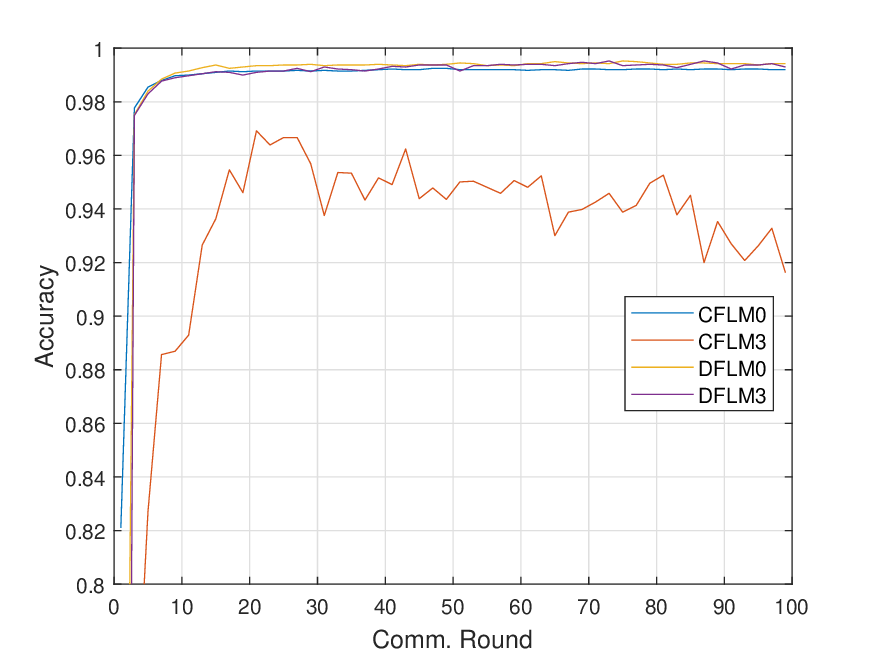}
    \caption{Training process of CFL and DFLoc-BFC under malicious attacks.}
    \label{test1}
\end{figure}

\begin{figure}
    \centering
    \includegraphics[width=\linewidth]{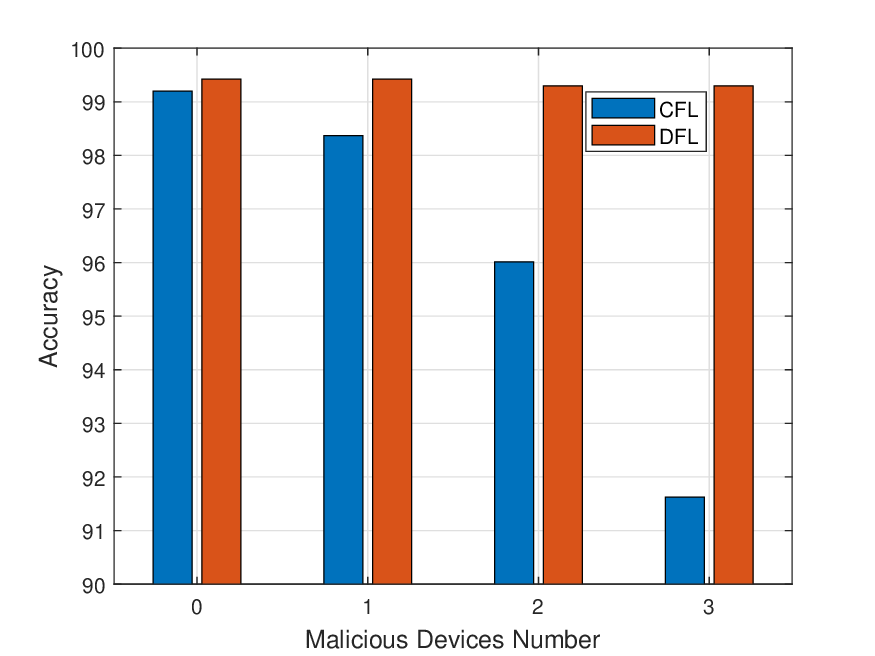}
    \caption{Effect of malicious attacks on CFL and DFLoc-BFC.}
    \label{test2}
\end{figure}

\subsubsection{Evaluation on the effect against malicious attacks on DFLoc-BFC}

We assess the classification accuracy of DFLoc-BFC, which measures the ratio of correct building-floor predictions to all predictions. Similar to the evaluation presented in Subsection~\ref{EofLLRMa}, a comparison of four schemes under a similar configuration is conducted, specifically CFLM0, CFLM3, DFLM0, and DFLM3 as shown in Fig.~\ref{test1}, and the results show similar trends. The orange curve denoting CFLM3 experiences difficulty in convergence, maintaining an accuracy plateau at approximately $90\%$, while the other three curves closely coincide. Fig.~\ref{test2} provides further insights into the vulnerability of CFL systems to malicious attacks, particularly when there are a notable number of malicious devices ($3$ malicious devices), resulting in a substantial accuracy decrease of $7.57\%$. Conversely, DFLoc-BFC consistently maintains a high accuracy rate, surpassing $99.2\%$.

\subsection{The Robustness Against Single-Point Failure}

To investigate the effectiveness of the proposed framework against single-point failure problems, we analyze the impact of the single-point failure on both the training and inference phases separately, considering a constant proportion of faulty devices in each working round. 

In the training phase of CFL, both servers and clients actively participate in the training process, server faults halt the entire system until the next round, while client faults result in parameter uploads with random values within the epoch. On the contrary, in the training phase of DFLoc, a faulty device during training, regardless of its role, is rendered inactive for the current round, due to its inability to pass signature checks during communication rounds. 


Within the inference phase of the CFL system, only the server remains active, and the outcome only depends on the server. Thus server faults during this phase result in a random output within the data range for the whole system. In contrast, in the inference phase of DFLoc, every client produces its respective outcome. Only malfunctioning devices generate random outputs, while the system takes into account each client's output to derive the final output. As a result, some post-processing procedures, such as aggregating client results (sorting and computing the median) can be employed in DFLoc to mitigate the impact of faulty devices and derive an expected final result.

\subsubsection{The Robustness Against Single-Point Failure on DFLoc-LLR}

\begin{figure}
    \centering
    \includegraphics[width=1\linewidth]{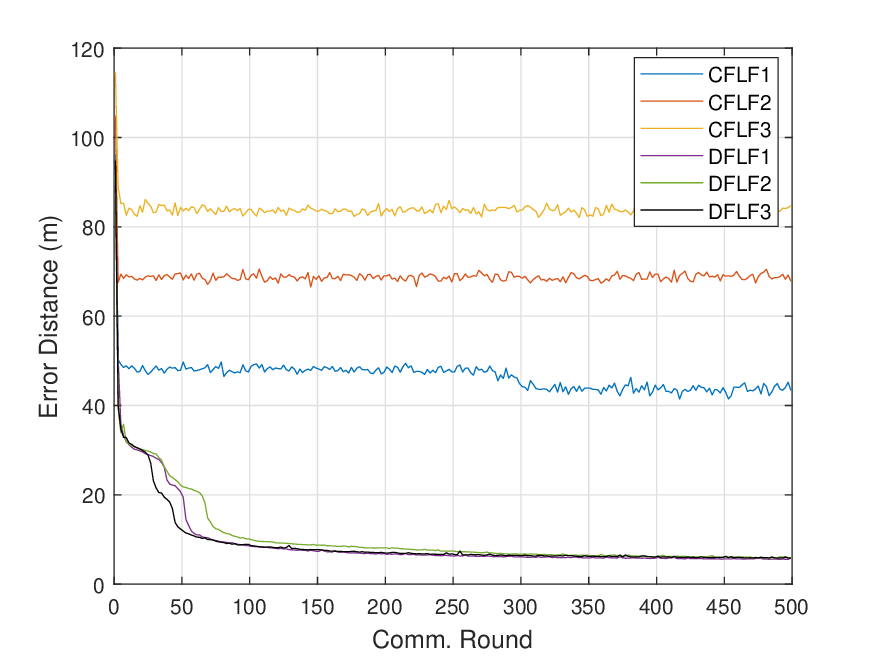}
    \caption{Training process of CFL and DFLoc-LLR under single-point failure.}
    \label{test7}
\end{figure}

This section investigates the impact of the single-point failure on the DFLoc-LLR during both the training and inference phases. Initially, experiments involving six schemes during the training phase are conducted: three for CFL and three for DFLoc, each with varying numbers of faulty devices in every training round. In Fig.~\ref{test7}, we observe that all the curves denoting CFL schemes (blue, orange, and yellow curves) show fluctuations. Furthermore, the curves denoting the DFLoc (purple, green, and black curves) exhibit rapid convergence and obvious overlap, indicating the framework's ability to mitigate the impact of single-point failure in the training phase. 

\begin{figure}
    \centering
    \includegraphics[width=1\linewidth]{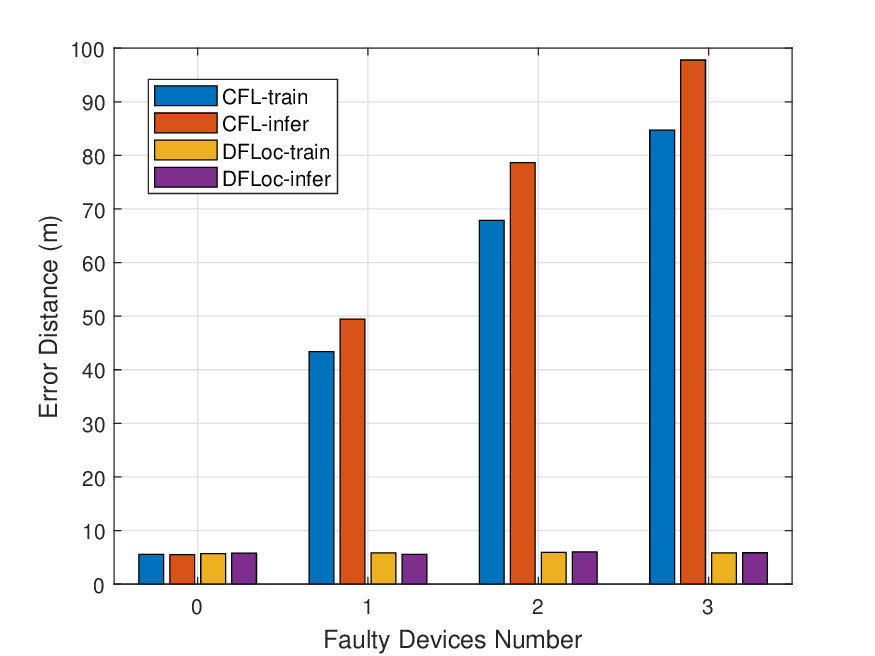}
    \caption{Effect of single-point failure on CFL and DFLoc-LLR.}
    \label{test8}
\end{figure}

Next, we analyze the inference phase of DFLoc-LLR through experiments conducted across four distinct groups, each corresponding to varying numbers of faulty devices in every round. Within each group, four schemes are assessed by the metric of error distance: CFL conducts the training phase only (blue bar), CFL conducts both training and inference phases (orange bar), and the same two schemes for DFLoc (yellow and purple bars). As shown in Fig.~\ref{test8}, when no faulty device is present, all four bars are identical. However, with an increasing number of faulty devices, rapid growth occurs in the first two values, while the latter two remain relatively stable. In a word, the DFLoc framework demonstrates robust resilience to the single-point failure in both phases, while traditional CFL systems are notably affected by such failures in these phases.

\subsubsection{The Robustness Against Single-Point Failure on DFLoc-BFC}

\begin{figure}
    \centering
    \includegraphics[width=1\linewidth]{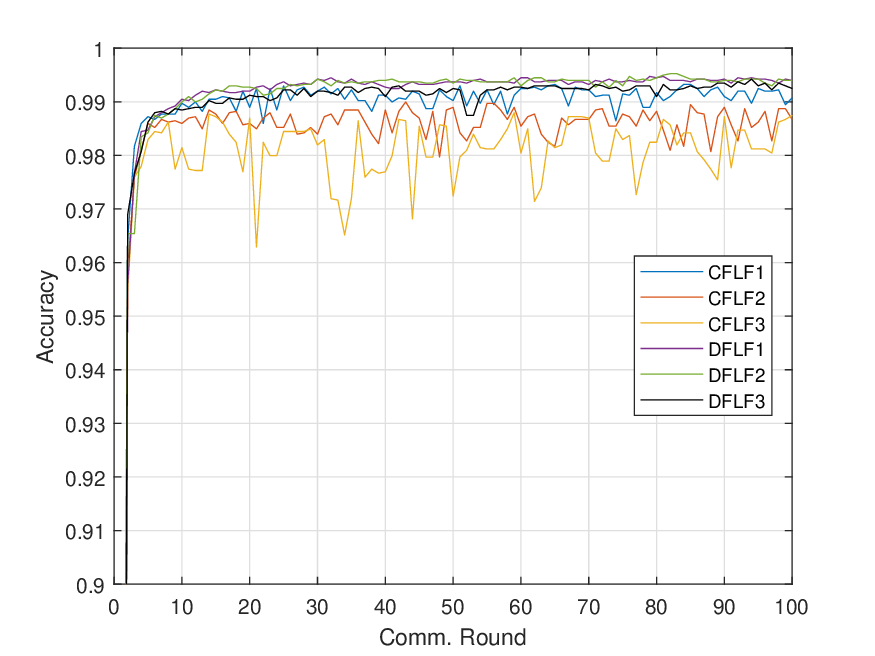}
    \caption{Training process of CFL and DFLoc-BFC under single-point failure.}
    \label{test3}
\end{figure}

\begin{figure}
    \centering
    \includegraphics[width=1\linewidth]{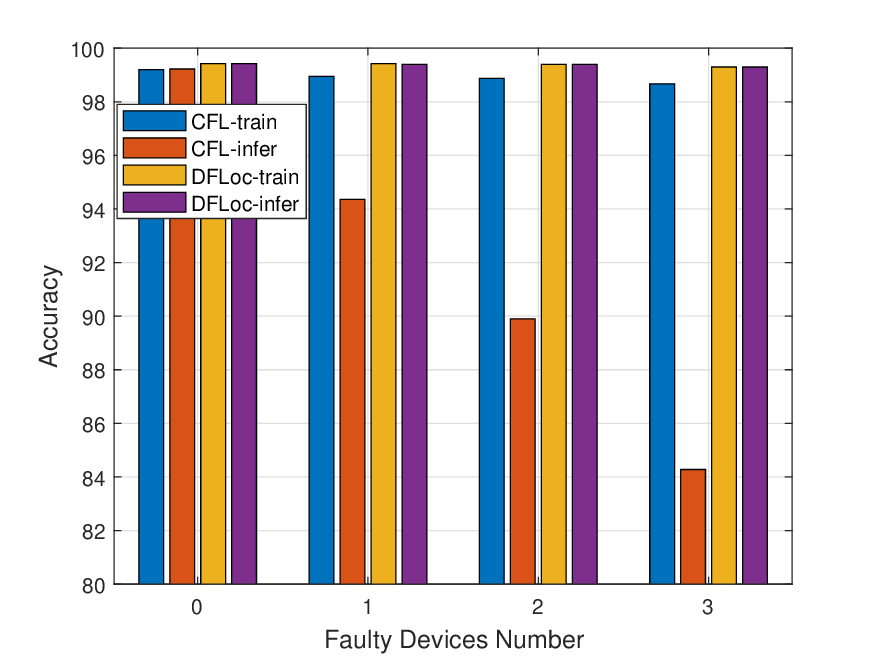}
    \caption{Effect of single-point failure on CFL and DFLoc-BFC.}
    \label{test4}
\end{figure}

In this section, the impact of the single-point failure on the DFLoc-BFC during both the training and inference phases is investigated. At first, a similar experiment is conducted with the same setup as the one shown in Fig.~\ref{test7}. In Fig.~\ref{test3}, we observe that the curves representing DFLoc (purple, green, and black curves) remain stable, while all curves belonging to CFL (blue, orange, and yellow curves) fail to converge. Then, an analogous experiment utilizing the identical configuration is conducted as depicted in Fig.~\ref{test8}. In Fig.~\ref{test4}, it is observed that as the number of faulty devices increases, the last two values within each group remain constant (yellow and purple bars), while the first two gradually decrease (blue and orange bars). This underscores the robust mitigation effect of our DFLoc framework against single-point failure. In contrast, traditional CFL systems show high susceptibility to single-point failure, with the main impact stemming from the inference phase within the DFLoc-BFC.

\begin{table*}[t]
\center
\label{table1}
\caption{Results for all experiments of CFL and DFL systems.}
\scalebox{1}{
\begin{tabular}{c|c|cccc|cccc|cccc}
\toprule

\multirow{2}{*}{\textbf{Concern}} & \multirow{2}{*}{\textbf{System}} & \multicolumn{4}{c|}{\textbf{BFC (accuracy /\%)}} & \multicolumn{4}{c|}{\textbf{LLR (error distance /m)}}    & \multicolumn{4}{c}{\textbf{3D (error distance /m)}}    \\

& & \textbf{0} & \textbf{1} & \textbf{2} & \textbf{3} & \textbf{0} & \textbf{1} & \textbf{2} & \textbf{3} & \textbf{0} & \textbf{1} & \textbf{2} & \textbf{3} \\
\midrule

 \multirow{2}{*}{Malicious Attack} 
 & CFL & 99.19     & 98.37     & 96.01     & 91.62   & 5.53      & 11.65     & 23.23     & 37.48  & 11.95 & 17.54 & 26.43 & 34.48  \\    & DFL  & 99.42     & 99.42     & 99.29     & 99.29  & 5.69      & 6.74      & 6.74      & 5.81 & 12.12 & 11.92 & 12.01 & 12.03\\
 \midrule
 
\multirow{2}{*}{Single-point Failure} 
& CFL & 99.22     & 94.35     & 89.89     & 84.27   & 5.49      & 49.43     & 78.66     & 97.79  & 12.24 & 50.85 & 79.91 & 98.64 \\    & DFL  & 99.42     & 99.39     & 99.39     & 99.29  & 5.79      & 5.54      & 6.03      & 5.85 & 12.39 & 12.26 & 12.52 & 12.41 \\
\midrule

\end{tabular}
}
\end{table*}

\subsection{Evaluation of DFLoc in 3D scenes} \label{3D_experiment}

Our framework is tailored for vast 3D scenes, including floor classification and latitude-longitude estimation. To accurately validate the overall performance of our DFLoc framework, we unify discrete building-floor indexes and continuous latitude-longitude values into a unified set of altitude, latitude, and longitude coordinates with consistent units. Since the UJIIndoorLoc dataset lacks explicit height data, we estimate height values based on floor information, assuming an average height of $6$ meters per floor.
Consequently, the BFC model enables us to obtain the height value for any location point, while the LLR model provides latitude and longitude values. Through the standardization of output labels, we can effectively evaluate the performance of our proposed DFLoc framework in 3D scenes under malicious attacks and single-point failure.

\begin{figure}[t]
    \centering
    \includegraphics[width=1\linewidth]{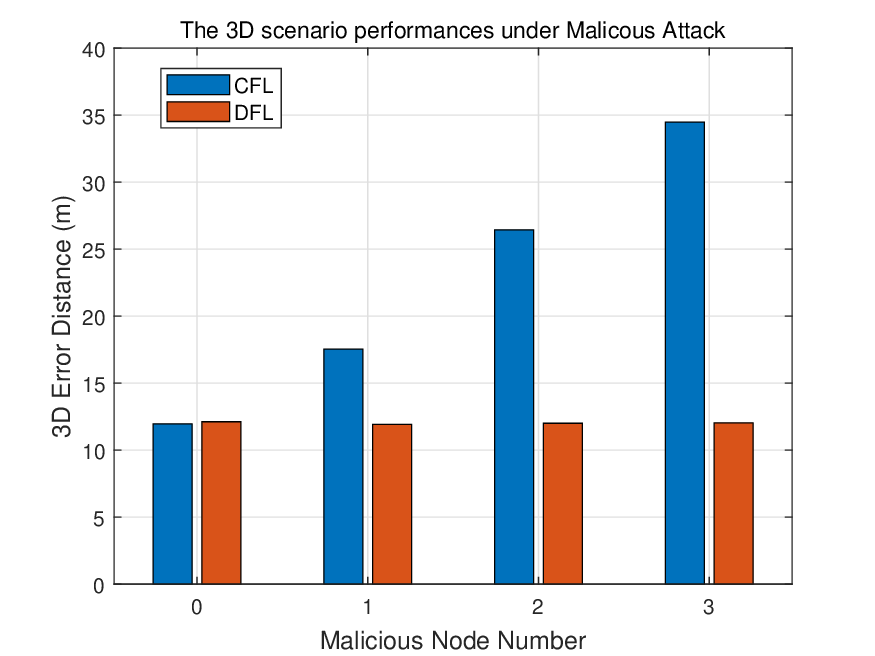}
    \caption{Performance of CFL and DFLoc-3D under malicious attacks.}
    \label{test9}
\end{figure}

First, an experiment for our DFLoc and traditional CFL in the 3D scene under malicious device attack is conducted as shown in Fig.~\ref{test9}, which demonstrates the 3D error distance of CFL and DFLoc under conditions of different numbers of malicious devices. When there are no malicious devices, the 3D error distances of CFL and DFLoc are very similar ($11.95$ m and $12.12$ m). With the increase in the number of malicious devices, the 3D error distance of CFL increases rapidly, while the DFLoc will remain stable. When the number of malicious devices increases to $3$, the error distance of both of them reaches $34.48$ m and $12.03$ m, respectively. This demonstrates that our DFLoc has a strong resistance capability against malicious attacks in the 3D environment, while the traditional CFL system is significantly influenced.

\begin{figure}
    \centering
    \includegraphics[width=1\linewidth]{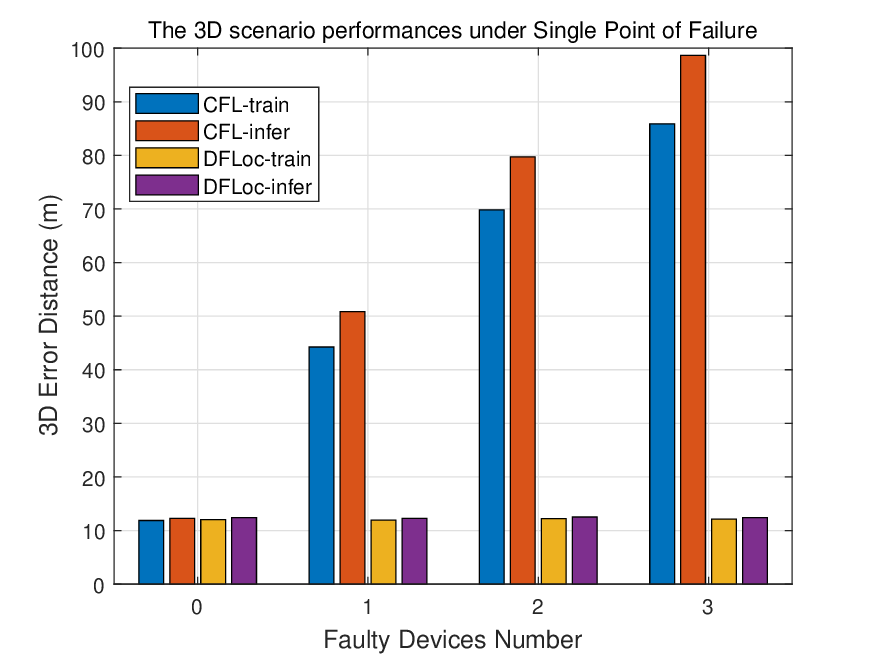}
    \caption{Performance of CFL and DFLoc-3D under single-point failure.}
    \label{test10}
\end{figure}

Then, the impact of DFLoc against the single-point failure is explored through an experiment with the same setup as the one shown in Fig.~\ref{test4}. In Fig.~\ref{test10}, it is evident that when there is no faulty device, the four values are nearly identical. However, as the number of faulty devices increases, rapid growth occurs in the blue and orange bars, while the yellow and purple bars remain constant. When the number of faulty devices reaches $3$, a significant gap becomes evident between the two bars denoting CFL and the other two bars denoting DFLoc. Both the yellow and purple bars stay at a small level of 3D error distance ($12.11$ m and $12.41$ m) while the blue and orange bars reach $85.87$ m and $98.64$ m. This observation demonstrates the superior performance of DFLoc in mitigating the impact of single-point failure compared to CFL in both phases.

\section{Conclusion} \label{V}

In this work, a novel framework named DFLoc is proposed to address the 3D indoor localization problem in large-scale and complex 3D indoor spaces, particularly under the challenges of malicious attacks and single-point failure. Within this framework, a decentralized federated learning approach is applied to train two networks: DFLoc-BFC and DFLoc-LLR, which provide accurate 3D position estimates based on RSS data. To bolster security and reliability, a model update validation mechanism and a decentralized architecture are incorporated into the federated learning process, effectively preventing the effects of malicious attacks and single-point failure. To better evaluate our DFLoc framework, we conduct extensive experiments using a real-world dataset of WiFi fingerprints. Our results demonstrate that DFLoc can effectively mitigate the challenges brought by malicious attacks and single-point failure in 3D environments when compared with the traditional central federated learning system. 

In future work, we will implement and evaluate DFLoc via real-world AIoT systems and blockchain data networks. Crowdsourced data will be collected from WiFi systems in smart buildings and subsequently annotated by users via a decentralized platform such as PublicAI. These labeled IoT data will be processed and stored on-chain within the Chainbase omni-chain data network. The localization functionality will be powered by the proposed DFLoc system, with location results translated into human-readable language by a crypto-native language model developed by Chainbase Inc. Through this approach, DFLoc aims to provide reliable data collection, annotation, and location-based services.

\bibliographystyle{unsrt}
\bibliography{ref}

\end{document}